\documentstyle[prc,aps,epsf]{revtex}
\begin{document}

\title{ New Observables in Break-up Reactions of 
        Three-body Halos. }
\author{E.~Garrido, D.V.~Fedorov and A.S.~Jensen \\
Institute of Physics and Astronomy, \\ Aarhus University, 
DK-8000 Aarhus C, Denmark}
\date{\today}

\maketitle

\begin{abstract}

Momentum distributions of particles from fast three-body halo
fragmentation reactions with polarized beams and outgoing fragments are
analyzed by use of the sudden approximation. The final state
interaction between the two non-disturbed particles is considered. We
first give a general and detailed description of the method. We
introduce observable quantities that emphasize the two-body
correlations in the initial three-body structure as well as in the
final two-body system. Using neutron removal from $^{11}$Li as an
example we investigate the dependence of the polarization observables
on the properties of the low-lying resonances and virtual states of the
$^{10}$Li subsystem. These observables are very sensitive to the 
$\ell > 0$-waves in the wave functions, and they provide detailed
information on the resonance structure of the neutron-core subsystem.

\end{abstract}
\vspace{5mm}
\par\leavevmode\hbox {\it PACS:\ } 25.60.Gc, 25.60.-t, 24.70.+s,
                                   21.45.+v
\vspace{5mm}

\section{Introduction} 

   Halo nuclei have been extensively investigated in the latest years
both from the theoretical and experimental point of view. For a review
see for instance \cite{zhuk93,bert93,han95} and references therein.

   The general properties of light nuclear halos have succesfully been
revealed by means of few-body models, dividing the degrees of freedom
into the approximately frozen (core) and active (halo).
Special attention has been paid to three-body Borromean bound
systems \cite{han87,joh90,fed94}, where none of the two-body
subsystems is bound. The most prominant examples are $^6$He
($^4$He+n+n) and $^{11}$Li ($^9$Li+n+n) both extensively discussed in
a general theoretical framework \cite{zhuk93}.

   The most detailed source of experimental information about the
structure of these nuclei is the momentum distributions of the
particles resulting from the fragmentation reactions
\cite{kob88,anne90,orr92,orr95,zin95,nils95,humb95}. For one particle
removal reaction with a high energy beam we can consider that the
removed particle in the halo projectile is instantaneously captured by the
target without disturbing the remaining two. This reaction picture is
known as sudden approximation, and has been proved to be fairly
accurate in the description of such halo fragmentation reactions
\cite{zhuk94,kors94,thom94,zhuk95,bar93,gar96}.  For a correct
interpretation of the experimental data it is essential to include the
interaction between the two non-disturbed particles after the
fragmentation, especially when low-lying resonances are present
\cite{zin95,kors94,bar93}. Detailed and consistent incorporation of
this Final State Interaction (FSI) into the model is necessary as seen
by its influence on the momentum distributions \cite{gar96}. Neutron
momentum distributions are very much affected by the final state
interaction, but also core momentum distributions are significantly
influenced.

   A polarized beam interacting with a target is not equally sensitive
to all the components of the wave function. It is then possible by
means of this tool to observe signals of the less dominant components,
and in this way to investigate finer details of the nuclear structure
of the projectile. The purpose of this paper is to investigate the
effects of polarized beams in fragmentation reactions of fast
three-body borromean halo nuclei. The case where the spin projection
of one of the particles is measured in the final state will also be
considered.  We introduce several useful observables and show the kind
of information we can extract from them. In particular we consider
neutron removal from the polarized $^{11}$Li projectile.  The
resulting neutron momentum distributions are sensitive to the amount
of $p$-state admixture in the neutron-core relative state.

The correct description of the structure and reactions of three-body
halo nuclei is an indispensable prerequisite for the present
investigation. These general procedures are discussed in previous
papers \cite{fed94,fed94b,gar95,cob96} which we shall use extensively.
This paper is organized as follows: In section II we formulate the
theoretical model and the method used to describe the fragmentation
reaction including definition of the polarization observables. The
numerical results obtained for $^{11}$Li are reported in section III
and in section IV we give a summary and the conclusions. We collect the
rather complicated analytic expressions of the differential momentum
distributions for different reactions in appendices A and B.

\section{Method}

The fragmentation reaction is described in the sudden approximation,
which assumes that one of the three particles in the projectile is
instantaneously removed by the target without disturbing the remaining
two. This approximation is valid for a high-energy three-body halo
beam interacting with a small target through short range interactions.
Therefore we assume a light target where the Coulomb dissociation
process only contributes marginally.

The transition matrix in the sudden approximation is given as 
\begin{equation} \label{tran} 
 M = \langle \Phi | \Psi \rangle \; , 
\end{equation} 
where $ \Phi$ is the final state wave function after break-up and
$\Psi$ is the initial wave function of the three-body projectile. We
now must specify these wave functions, compute the matrix element and
integrate $|M|^2$ over all the non observed variables.

\subsection{Wave functions}

The initial wave function describes the bound three-body halo system
whereas the final state wave function describes the two-body
continuum state specified by the experimental conditions.

\subsubsection{Initial wave function}

The projectile three-body wave function $\Psi$ with inclusion of the
proper quantum numbers and variables is denoted by $\Psi^{J M}(\bbox{x}, 
\bbox{y})$, where $J$ and $M$ are the total spin and its
projection.  The coordinates $\bbox{x}$ and $\bbox{y}$ are the usual
Jacobi coordinates \cite{fed94b}, where $\bbox{x}$ is drawn between
the two particles surviving after the fragmentation. The wave function
is expanded in terms of a complete set of hyperspherical harmonics
$\bbox{Y}_{\ell_x \ell_y}^{K L}(\alpha, \Omega_{x}, \Omega_{y})$,
where the quantum number $K$ usually is called the hypermoment,
$\ell_x$ and $\ell_y$ are the orbital angular momenta associated with
$\bbox{x}$ and $\bbox{y}$, and $L$ is resulting from the coupling of
these angular momenta.  The variables $\rho$, $\alpha$, $\Omega_x$,
and $\Omega_y$ are the hyperspherical variables obtained from the
Jacobi coordinates $\bbox{x}$ and $\bbox{y}$ \cite{fed94}. Then $\Psi^{J
M}(\bbox{x}, \bbox{y})$ is written as
\begin{eqnarray} 
\lefteqn{\Psi^{J M}(\bbox{x}, \bbox{y}) =} \nonumber \\
& &
\frac{1}{\rho^{5/2}} \sum_n f_n(\rho) \sum_{K \ell_x \ell_y L s_x S}
C_{n K \ell_x \ell_y L s_x S}(\rho) \left[ \bbox{Y}_{\ell_x
\ell_y}^{K L}(\alpha, \Omega_x, \Omega_y) \otimes \chi_{s_x s_y}^S
\right]^{J M}  \; , 
\label{totrot} 
\end{eqnarray} 
where the spin function $\chi_{s_x s_y}^S$ is obtained by coupling the
spins of two of the particles to $s_x$ which in turn is coupled to the
spin $s_y$ of the third particle resulting in the total spin $S$.  The
radial functions $f_n(\rho)$ and the expansion coefficients $C_{n K
\ell_x \ell_y L s_x S}(\rho)$ are calculated numerically by solving
the Faddeev equations in coordinate space \cite{fed94b}.  Details
about this procedure can be found in \cite{fed94b}.

\subsubsection{Final state wave function}

The final state wave function $\Phi$ in the plane wave approximation in
the center of mass frame of the projectile is given by 
\begin{equation} \label{e5}
   \Phi =  e^{i \mbox{\small $\bbox{k}$} \cdot \mbox{\small $\bbox{R}$} }
   \chi^{\sigma_y}_{s_y} 
 e^{i \mbox{\small $\bbox{q}$} \cdot \mbox{\small $\bbox{r}$} }
 \chi^{\sigma_1}_{s_1}   \chi^{\sigma_2}_{s_2}    \; ,
\end{equation}
where $ \chi^{\sigma_i}_{s_i}$ for i=1,2 and $\chi^{\sigma_y}_{s_y}$
are the spin wave functions respectively for the two undisturbed
particles and the third removed particle. Here $s$ and $\sigma$ are
the total spin and its projection on an axis. Furthermore $\bbox{k}$
and $\bbox{q}$ are the total and relative momentum of the two
remaining particles in the final state, $\bbox{r}$ is the distance
between the two remaining particles, and $\bbox{R}$ is the distance
between the center of mass of this two-body system and the removed
particle.

The description in terms of plane waves in Eq.(\ref{e5}) implies that
all the interactions between the particles are neglected in the final
state wave function. However, in our approximation only one of the
particles is removed from the other two, and the interaction
between the non-disturbed particles can not be neglected.  In a recent
work \cite{gar96} we investigated the influence of this final state
interaction (FSI) on the momentum distributions, and observed that
they are crucial for reproduction of the experimental momentum
distributions.

  In Eq.(\ref{e5}) introduction of the final state interaction amounts
to substitution of the plane wave, $ e^{i \mbox{\small $\bbox{q}$}
\cdot \mbox{\small $\bbox{r}$} } \chi^{\sigma_1}_{s_1}
\chi^{\sigma_2}_{s_2}$, describing the two-body system in the final
state by the appropriate {\it distorted} two-body wave function
$w$. When the two-body interaction do not mix the two-body states with
different spin $s_x$ and relative orbital angular momentum $\ell_x$,
we can expand this two-body wave function ($w$) in partial waves as in
\cite{newt82}. These assumptions are usually strictly valid. The only
exception arises from the tensor interaction. However the
resulting mixing is often very small and therefore insignificant in
the present context. We then have
\begin{equation}  \label{2bexp}
w^{\sigma_1 \sigma_2}(\bbox{k}_x, \bbox{x}) =
\sum_{s_x \sigma_x} \langle s_1  \sigma_1 s_2 \sigma_2|s_x \sigma_x \rangle
w^{s_x \sigma_x}(\bbox{k}_x, \bbox{x}),
\end{equation}
with
\begin{eqnarray}\label{e7}
w^{s_x \sigma_x}(\bbox{k}_x, \bbox{x})= \sqrt{\frac{2}{\pi}}
\frac{1}{k_x x} \sum_{j_x \ell_x m_x} u_{\ell_x s_x}^{j_x} (k_x,x)
{\cal Y}_{j_x \ell_x s_x}^{m^*_x}(\Omega_x)
 \nonumber \\
\times \sum_{m_{\ell_x}=-\ell_x}^{\ell_x}
\langle \ell_x m_{\ell_x} s_x \sigma_x|j_x m_x \rangle i^{\ell_x}
                        Y_{\ell_x m_{\ell_x}}(\Omega_{k_x})  \; ,
\end{eqnarray}
where $\bbox{k}_x$ is the momentum associated with $\bbox{x}$,
$\sigma_x$ is the spin projection of $s_x$, $u$ is the radial and
${\cal Y}$ the angular distorted wave function. The angles $\Omega_x$
and $\Omega_{k_x}$ define the direction of $\bbox{x}$ and
$\bbox{k}_x$, respectively.  The dependence on the spins $s_1$ and
$s_2$ of the two particles surviving after the fragmentation reaction
has been omitted for notational simplicity.

The radial wave functions $u_{\ell_x s_x}^{j_x}(k_x,x)$ are obtained
numerically
by solving the Schr\"odinger equation with the appropriate two-body
potential
$\hat{V}(\bbox{x})$
\begin{equation}
\frac{\partial^2}{\partial x^2} u_{\ell_x s_x}^{j_x}(k_x,x)+
\left( k_x^2 - \frac{2 m}{\hbar^2} V_{\ell_x s_x}^{j_x}(x)
- \frac{\ell_x (\ell_x+1)}{x^2} \right)
u_{\ell_x s_x}^{j_x}(k_x,x)=0  \; ,
\label{schr}
\end{equation}
where the effective radial potential is obtained by integration over
the angles $\Omega_x$
\begin{equation}
V_{\ell_x s_x}^{j_x}(x)=\int d\Omega_x
                        {\cal Y}_{j_x \ell_x s_x}^{m_x^*}(\Omega_x)
  \hat{V}(\bbox{x}) {\cal Y}_{j_x \ell_x s_x}^{m_x}(\Omega_x)  \; 
\end{equation}
and $m$ is an arbitrary normalization mass.

\subsection{Transition matrix element}

The calculation of the transition matrix requires now the overlap of
the initial three-body wave function in Eq.(\ref{totrot}) and the
final state wave function $e^{i \mbox{\small $\bbox{k}$} \cdot
\mbox{\small $\bbox{R}$} } \chi_{s_y}^{\sigma_y} w^{\sigma_1
\sigma_2}(\bbox{k}_x, \bbox{x})$, where $\bbox{k}_y$ is the momentum
associated with the $\bbox{y}$ coordinate. Since $\bbox{k}_y \cdot
\bbox{y} = \bbox{k} \cdot \bbox{R}$ we then find
\begin{equation}
M^{J M}_{\sigma_1 \sigma_2 \sigma_y}(\bbox{k}_x, \bbox{k}_y) =
\sum_{s_x \sigma_x} \langle s_1 \sigma_1 s_2 \sigma_2|s_x \sigma_x \rangle
M^{J M}_{s_x \sigma_x s_y \sigma_y}(\bbox{k}_x, \bbox{k}_y) 
\label{tm}
\end{equation}
For simplicity we do not specify the dependence of
$M^{J M}_{\sigma_1 \sigma_2 \sigma_y}(\bbox{k}_x, \bbox{k}_y)$
on the spins $s_1$, $s_2$ and $s_y$ of the three particles involved in
the reaction.

The analytic form of $M^{J M}_{s_x \sigma_x s_y \sigma_y}(\bbox{k}_x,
\bbox{k}_y)$ is
\begin{eqnarray} 
M^{J M}_{s_x \sigma_x s_y \sigma_y}(\bbox{k}_x,
\bbox{k}_y) & \propto & \frac{2}{\pi} \sum_{\ell_x m_{\ell_x}
\ell_y m_{\ell_y}} \sum_{j_x L S} I_{\ell_x s_x j_x}^{\ell_y L
S}(\kappa, \alpha_\kappa) Y_{\ell_x m_{\ell_x}}(\Omega_{k_x}) Y_{\ell_y
m_{\ell_y}}(\Omega_{k_y})
    \nonumber \\ 
& & \hspace{-2cm} \times \sum_{m_x j_y m_y}
(-1)^{J+2S-2M+\ell_y+s_y-s_x-\ell_x} \hat{j}_x^2 \hat{j}_y^2 \hat{J}
\hat{L} \hat{S} \left(
    \begin{array}{ccc}
       J & j_x & j_y \\ M & -m_x & -m_y \end{array} \right)
	       \nonumber \\ 
& & \hspace{-2cm} \times
 \left(
    \begin{array}{ccc}
       j_y & \ell_y & s_y \\ -m_y & m_{\ell_y} & \sigma_y \end{array}
    \right) \left(
    \begin{array}{ccc}
       j_x & \ell_x & s_x \\ -m_x & m_{\ell_x} & \sigma_x \end{array}
    \right) \left\{
    \begin{array}{ccc}
       J & j_x & j_y \\ L & \ell_x & \ell_y \\ S & s_x  & s_y
    \end{array} \right\} 
\label{tm0} 
\end{eqnarray} 
where $\hat{a}=\sqrt{2a+1}$. In Eq.(\ref{tm0}) we
have introduced the hyperspherical coordinates in momentum space
$\kappa=\sqrt{k_x^2+k_y^2}$, $\alpha_\kappa=\arctan{(k_x/k_y)}$, and
the numerical function $I^{\ell_y L S}_{\ell_x s_x j_x}(\kappa,
\alpha_\kappa)$ is given by 
\begin{eqnarray} 
I^{\ell_y L S}_{\ell_x s_x j_x}(\kappa, \alpha_\kappa)& = & 
           i^{\ell_x+\ell_y} \sum_{K n} N_K^{\ell_x \ell_y} 
   \int \rho^{5/2} d\rho f_n(\rho) C_{n K \ell_x \ell_y L s_x S}(\rho)
	 \nonumber \\ & &  \hspace{-3cm}
\left[ \int d\alpha (\sin \alpha)^{\ell_x+2} (\cos \alpha)^{\ell_y+2}
P_\nu^{\ell_x+\frac{1}{2},\ell_y+\frac{1}{2}}(\cos(2\alpha))
j_{\ell_y}(k_y y) u_{\ell_x s_x}^{j_x}(k_x, x) \right] 
\label{ifun}
\end{eqnarray} 
where $\nu=(K-\ell_x-\ell_y)/2$, $P_\nu^{\ell_x+\frac{1}{2},
\ell_y+\frac{1}{2}}$ is a Jacobi polynomial, and 
\begin{equation}
N_K^{\ell_x \ell_y}= \left[ \frac{ \nu! (\nu+\ell_x+\ell_y+1)!2(K+2) }
	{\Gamma(\nu+\ell_x+\frac{3}{2}) \Gamma(\nu+\ell_y+\frac{3}{2})}
		\right]^{1/2}.
\end{equation}

\subsection{Cross sections}

The cross section or momentum distribution is now obtained by squaring
the transition matrix, and subsequently averaging over initial states
and summing up over final states. If we consider the possibility of
having a polarized beam and measuring the spin projection of the two
particles in the final state we can then write
\begin{equation}
\frac{d^6\sigma}{d\bbox{k}_x d\bbox{k}_y} \propto
\sum_M W_{\mbox{\scriptsize init.}}(M) \sum_{\sigma_1} 
W_{\mbox{\scriptsize fin.}}(\sigma_1) \sum_{\sigma_2}
W_{\mbox{\scriptsize fin.}}(\sigma_2) \sum_{\sigma_y}
|M^{J M}_{\sigma_1 \sigma_2 \sigma_y}(\bbox{k}_x, \bbox{k}_y)|^2
\label{mom}
\end{equation}
where $W_{\mbox{\scriptsize init.}}(M)$, $W_{\mbox{\scriptsize
fin.}}(\sigma_1)$, and $W_{\mbox{\scriptsize fin.}}(\sigma_2)$ give
the occupation probability of each magnetic substate in the initial
and final states. When the beam is unpolarized and there is no
detection of final spin projections these probabilities take the
values $W_{\mbox{\scriptsize fin.}}(M)=1/(2J+1)$,
$W_{\mbox{\scriptsize fin.}}(\sigma_1)=1/(2s_1+1)$, and
$W_{\mbox{\scriptsize fin.}}(\sigma_2)=1/(2s_2+1)$, and we obtain the
unpolarized expression for the momentum distributions \cite{gar96}.

The volume element $d\bbox{k}_x d\bbox{k}_y$ is written 
\begin{equation}
d\bbox{k}_x d\bbox{k}_y=
k_x^\bot dk_x^\bot dk_x^\parallel d\varphi_{k_x} k_y^2 dk_y d\Omega_{k_y}
\end{equation}
where $k_x^\bot = k_x \sin{\theta_{k_x}}$, $k_x^\parallel=k_x
\cos{\theta_{k_x}}$, and $(\theta_{k_x}, \varphi_{k_x})$ are the polar
and azimuthal angles associated to $\Omega_{k_x}$.

The integration of Eq.(\ref{mom}) over $\Omega_{k_y}$ and
$\varphi_{k_x}$ can be done analytically.  By further numerical
integration over $k_y$ and $k_x^\bot$, we get the
one-dimensional relative momentum $(k_x^\parallel)$ distribution of
the remaining particles.  By integrating over $k_y$ and
$k_x^\parallel$, we get instead the two-dimensional relative momentum
($k_x^\bot$) distribution.  It should be noted that we have not
specified any coordinate system and the axes $\bbox{x}$, $\bbox{y}$,
and $\bbox{z}$ are therefore completely arbitrary.  Thus, in the
sudden approximation the longitudinal and transverse momentum
distributions are identical.

After substituting Eqs.(\ref{tm}) and (\ref{tm0}) into (\ref{mom}), and
integrating analytically over $\Omega_{k_y}$ and $\varphi_{k_x}$ we
obtain the expressions for the three-differential momentum
distributions that are collected in appendix A. Three different cases
are considered: 

(i) Eq.(\ref{rel1}) describes the process with 100\% polarized beam
where the polarization of one of the particles in the final state is
measured, i.e. $W_{\mbox{\scriptsize init.}}(M^\prime)=\delta_{M
M^\prime}$, $W_{\mbox{\scriptsize fin.}}(\sigma_1^\prime)=
\delta_{\sigma_1 \sigma_1^\prime}$, $W_{\mbox{\scriptsize
fin.}}(\sigma_2^\prime)=1/(2s_2+1)$,

(ii) Eq.(\ref{rel2}) describes the process with 100\% polarized beam
without detection of the final spin states, i.e. $W_{\mbox{\scriptsize
init.}}(M^\prime)=\delta_{M M^\prime}$, $W_{\mbox{\scriptsize
fin.}}(\sigma_1^\prime)=1/(2s_1+1)$, $W_{\mbox{\scriptsize
fin.}}(\sigma_2^\prime)=1/(2s_2+1)$,

(iii) Eq.(\ref{rel3}) describes the process with unpolarized beam
without detection of the final spin states, i.e. $W_{\mbox{\scriptsize
init.}}(M^\prime)=1/(2J+1)$, $W_{\mbox{\scriptsize
fin.}}(\sigma_1^\prime)=1/(2s_1+1)$, $W_{\mbox{\scriptsize
fin.}}(\sigma_2^\prime)=1/(2s_2+1)$.

To compare with the experimental data the momentum distributions
should be referred to the center of mass frame of the projectile. To
do this we construct the momentum $\bbox{p}$ of one of the particles
in the final state relative to the center of mass of the projectile as
a linear combination of $\bbox{k}_x$ and $\bbox{k}_y$, i.e.
\begin{equation}
\bbox{p} = a_i \bbox{k}_x + b_i \bbox{k}_y \; ,
\label{mcm}
\end{equation}
where $i=1,2$ refer to the two remaining particles.  If the Jacobi
coordinate $\bbox{x}$ goes from particle 1 to particle 2, then $a$ and
$b$ take the values
\begin{equation}
a_1=-\left( \frac{1}{m} \frac{m_1 m_2}{m_1+m_2} \right)^{1/2},
\hspace{0.5cm}
b_1=\frac{m_1}{m_1+m_2} \left(\frac{1}{m}
             \frac{(m_1+m_2)m_3}{m_1+m_2+m_3} \right)^{1/2} \; ,
\label{ab1}
\end{equation}

\begin{equation}
a_2= \left( \frac{1}{m} \frac{m_1 m_2}{m_1+m_2} \right)^{1/2},
\hspace{0.5cm}
b_2=\frac{m_2}{m_1+m_2} \left(\frac{1}{m}
             \frac{(m_1+m_2)m_3}{m_1+m_2+m_3} \right)^{1/2} \; ,
\label{ab2}
\end{equation}
when we compute the momentum distributions of particle 1 and 2,
respectively.

Since the volume element $d\bbox{p}d\bbox{k}_y = a_i^3 d\bbox{k}_x
d\bbox{k}_y$ we obtain the relation
\begin{equation}
\frac{d^6\sigma}{d\bbox{p} d\bbox{k}_y} = 
  \frac{1}{a_i^3} \frac{d^6\sigma}{d\bbox{k}_x d\bbox{k}_y} \; .
\label{rot}
\end{equation}
The only additional complication in this coordinate system is that the
numerical function $I^{\ell_y L S}_{\ell_x s_x j_x}(\kappa,
\alpha_\kappa)$ now depends on the relative angle
$\theta_{k_y}^\prime$ between $\bbox{p}$ and $\bbox{k}_y$. Only two
angular integrations can then be performed analytically, and three
integrations must be done numerically in order to get the
one-dimensional and two-dimensional momentum distributions of one of
the particles in the final state relative to the center of mass of the
projectile.

In appendix B we show the analytic expressions for the
four-differential momentum distributions relative to the center of
mass of the projectile for the same three cases as specified in appendix
A. They are given by Eqs.(\ref{cm1}), (\ref{cm2}) and (\ref{cm3}),
respectively.

\subsection{Polarization Observables}

We shall denote the one and two-dimensional momentum distributions
when the beam is unpolarized and the spin projection of the particles
is not measured in the final state by ${\cal Q}_1(p^\parallel) =
d\sigma/d p^\parallel$ and ${\cal Q}_2(p^\bot) = d\sigma/d p^\bot$,
respectively.  The distributions are referred to the center of mass of
the three-body projectile, and obtained by integration over the
unobserved quantities of Eq.(\ref{cm3}) in appendix B.

In the same way we shall denote the one and two-dimensional momentum
distributions when the beam is 100\% polarized (with spin projection
$M$) and the spin projection of the particles in the final state is
not measured by ${\cal Q}^M_1(p^\parallel) = d\sigma^M/d p^\parallel$
and ${\cal Q}^M_2(p^\bot)= d\sigma^M/d p^\bot$, respectively.  These
distributions are obtained by integration of Eqs.(\ref{cm2}) in
appendix B.

Finally, ${\cal Q}^{M \sigma}_1(p^\parallel)= d\sigma^{M \sigma}/d
p^\parallel$ and ${\cal Q}^{M \sigma}_2(p^\bot)= d\sigma^{M \sigma}/d
p^\bot$ will denote the one and two-dimensional momentum distributions
for a 100\% polarized beam (with spin projection $M$), and when the
spin projection of one of the particles in the final state is measured
to be $\sigma$.  These distributions are obtained by integration of
Eqs.(\ref{cm1}) in appendix B.

We define now the following asymmetries
\begin{equation}
A_i^{M M^\prime} \equiv
\frac{{\cal Q}_i^M - {\cal Q}_i^{M^\prime}}
     {{\cal Q}_i}, \hspace{0.7cm} i=1,2
\label{as1}
\end{equation}
\begin{equation}
A_i^{M M^\prime;\sigma \sigma^\prime} \equiv
\frac{ {\cal Q}_i^{M \sigma}
           - {\cal Q}_i^{M^\prime \sigma^\prime} }
     { {\cal Q}_i },  \hspace{0.7cm} i=1,2
\label{as2}
\end{equation}

It is easy to check that the $I=I^\prime=0$ term in Eq.(\ref{cm1}) and
the $I=0$ term in Eq.(\ref{cm2}) are independent of $M$ and $\sigma$,
and equal to the totally unpolarized momentum distribution
Eq.(\ref{cm3}). This means that the deviations from zero of the
asymmetries Eqs.(\ref{as1}) and (\ref{as2}) are a direct consequence of
the polarization of the beam and the measure of the spin projection of
one of the particles after the fragmentation.

Direct examination of Eq.(\ref{cm2}) permits us to observe that if
only $s$-waves were involved in the initial and final wave functions
($\ell_x=\ell^\prime_x=L_x=0$ and $\ell_y=\ell^\prime_y=L_y=0$) only
the $I$=0 term would contribute to the momentum distribution. In other
words, the asymmetry Eq.(\ref{as1}) would vanish.  Therefore, the
deviation from zero of this asymmetry is a signal of non-vanishing
components different from s-waves in the wave function. This
conclusion is not true for the asymmetry Eq.(\ref{as2}), since the
momentum distribution Eq.(\ref{cm1}) has additional non-vanishing
terms even when only $s$-waves are involved.

\begin{figure}[t]
\epsfxsize=12cm
\epsfysize=7cm
\epsfbox[650 200 1200 550]{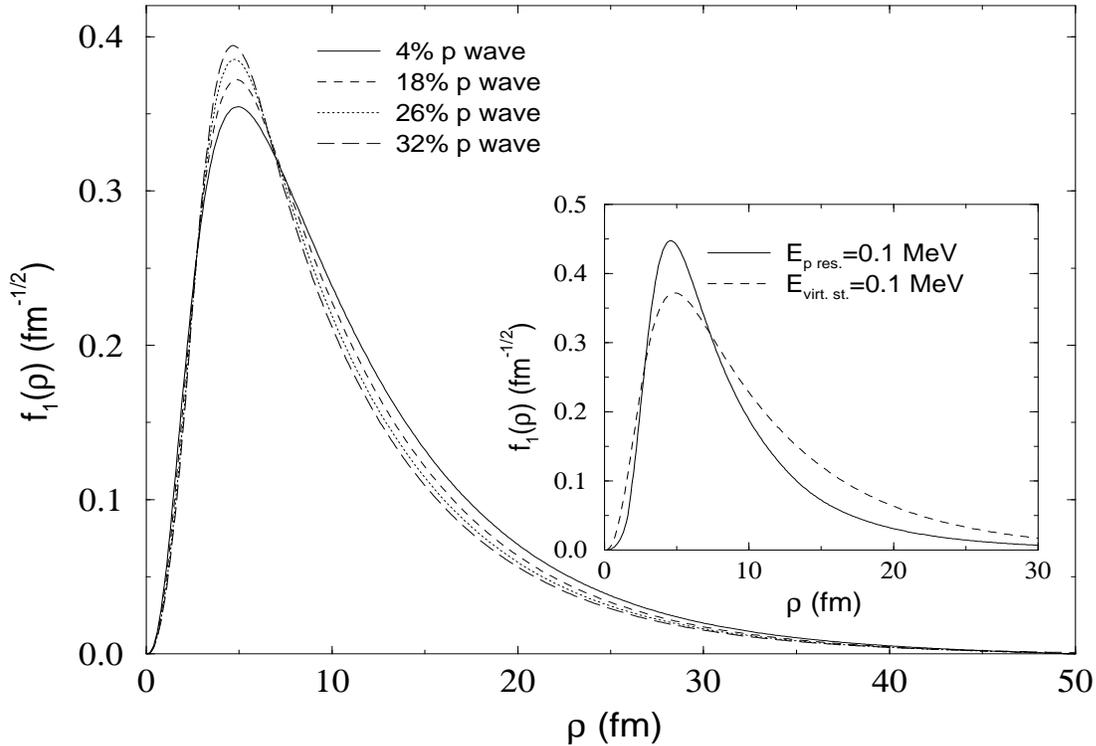}
\vspace{3.5cm}
\caption{\protect\small
The radial wave function $f_1(\rho)$ for $^{11}$Li when
$^{10}$Li has a virtual $s$-state at 100 keV the lowest-lying
$p$-resonance at 0.5 MeV. Calculations with 4\% of $p$-wave (solid
line), 18\% of $p$-wave (short-dashed line), 26\% of $p$-wave (dotted
line), and 32\% (long-dashed line) in the $^{10}$Li wave function are
shown. The inset shows the same wave function when $^{10}$Li has a
low-lying virtual $s$-state at 100 keV (dashed line) and when
$^{10}$Li has a low-lying $p$-resonance at 100 keV (solid line).
}
\label{1}
\end{figure}

\section{Results}

In this section we apply the procedure described above to two-neutron
halo nuclei, and in particular to $^{11}$Li. We consider neutron
removal reactions, and the neutron-$^9$Li interaction is include in
the final state. We only included $s$- and $p$-waves in the
calculations, since components with higher angular momenta are
negligibly small. We use the two-body potentials described in
ref.\cite{gar95}, i.e.
\begin{equation}
 V_{nn}=\left( V_c+V_{ss} \bbox{s}_{n1} \cdot \bbox{s}_{n2}
+V_T \hat{S}_{12}+V_{so} \bbox{l}_{nn} \cdot \bbox{s}_{nn}
\right) \exp \left[ - (r/b_{nn})^2 \right] 
\end{equation}
\begin{eqnarray}
V_{nc}^{(s)}&=&V_s (1+\gamma_s \mbox{\bf s}_c \cdot \mbox{\bf s}_n)
   \exp \left[ - (r/b_{nc})^2 \right], \label{pots} \\
V_{nc}^{(l)}&=&(V_l + V_{so}^{(l)} \mbox{\bf l}_{nc} \cdot \mbox{\bf
s}_{nc})
   \exp \left[ - (r/b_{nc})^2 \right] 
\label{potl}  \; ,
\end{eqnarray}

The nucleon-nucleon potential $V_{nn}$ is made of gaussians containing
central ($V_c$), spin splitting ($V_{ss}$), spin-orbit ($V_{so}$) and
tensor ($V_T$) parts where the parameters are fitted to low energy
$s$- and $p$-wave nucleon-nucleon scattering data ($\bbox{s}_{n1}$ and
$\bbox{s}_{n2}$ are the spins of the two neutrons, $\bbox{l}_{nn}$ is
the two-neutron relative orbital angular momentum,
$\bbox{s}_{nn}=\bbox{s}_{n1}+\bbox{s}_{n2}$, and $\hat{S}_{12}$ is the
usual tensor operator). In the neutron-neutron subsystem
more than 98\% of the probablity is found in $s$-waves.

The neutron-core potential $V_{nc}$ is also made of gaussians adjusted
to reproduce the experimental binding energy and root mean square
radius of the three-body projectile ($295 \pm 35$ keV \cite{you93} and
$3.1 \pm 0.3$ fm \cite{tani92}, respectively, for $^{11}$Li).  The
different waves are independently fitted, allowing us to adjust
separately the energy positions of the virtual $s$-states and the
resonances in the neutron-core subsystem. For simplicity the spin
splitting term (introduced by the parameter $\gamma_s$) has been
included only for $s$-waves in the neutron-core interaction.

\subsection{Initial and final wave functions}

The expression we use for the wave function of the three-body halo
projectile is given by Eq.(\ref{totrot}), where the radial functions
$f_n(\rho)$ and the coefficients $C_{n K \ell_x \ell_y L s_x S}(\rho)$
are computed numerically. To illustrate some of the general aspects of
the three-body wave functions, we show in Fig. \ref{1} the $f_1(\rho)$
function for $^{11}$Li. This function dominates the expansion in
Eq.(\ref{totrot}) as it contains more than 95\% of the norm of the
total wave function. The neutron-core interaction has been fitted to
produce a low-lying virtual $s$-state at 100 keV and the lowest
$p$-resonance at 0.5 MeV. By use of the spin-orbit parameter
($V_{so}^{(l=1)}$) we can move the higher-lying p-resonances up or down
and thereby modify the $p$-wave content in the neutron-$^9$Li
subsystem. Four different situations of varying $p$-wave content are
shown, i.e. 4\% (solid line), 18\% (short-dashed line), 26\% (dotted
line) and 32\% (long-dashed line). We observe that the more $p$-wave
content in the neutron-$^9$Li subsystem the narrower the radial wave
function. This result can be attributed to the centrifugal barrier,
that keeps the neutrons closer to the core for $p$-waves.

When the final state interaction is neglected in the calculation, the
momentum distributions are given by the square of the Fourier
transform of the three-body wave function in
Eq.(\ref{totrot}). Therefore a larger $p$-wave content in the
neutron-core subsystem will produce a broader momentum
distribution. Details about calculations of momentum distributions
following this method have been presented in \cite{gar96}.
Nevertheless, the differences between the three-body wave functions in
Fig. \ref{1} are not very large and the corresponding momentum distributions
would also be very similar. In the inset of Fig. \ref{1} we display the same
radial three-body wave function $f_1(\rho)$ for a neutron-core
potential producing a $p$-resonance at 100 keV (solid line) and a
virtual $s$-state at 100 keV (dashed line). Again we find that the
low-lying $p$-resonance produces a narrower radial wave function, and
therefore a broader momentum distribution (without FSI). In this case
the difference between the wave functions is larger and as a
consequence the momentum distributions for a low-lying $p$-resonance
and a low-lying virtual $s$-state will also be clearly
distinguishable.

\begin{figure}[t]
\epsfxsize=12cm
\epsfysize=7cm
\epsfbox[650 200 1200 550]{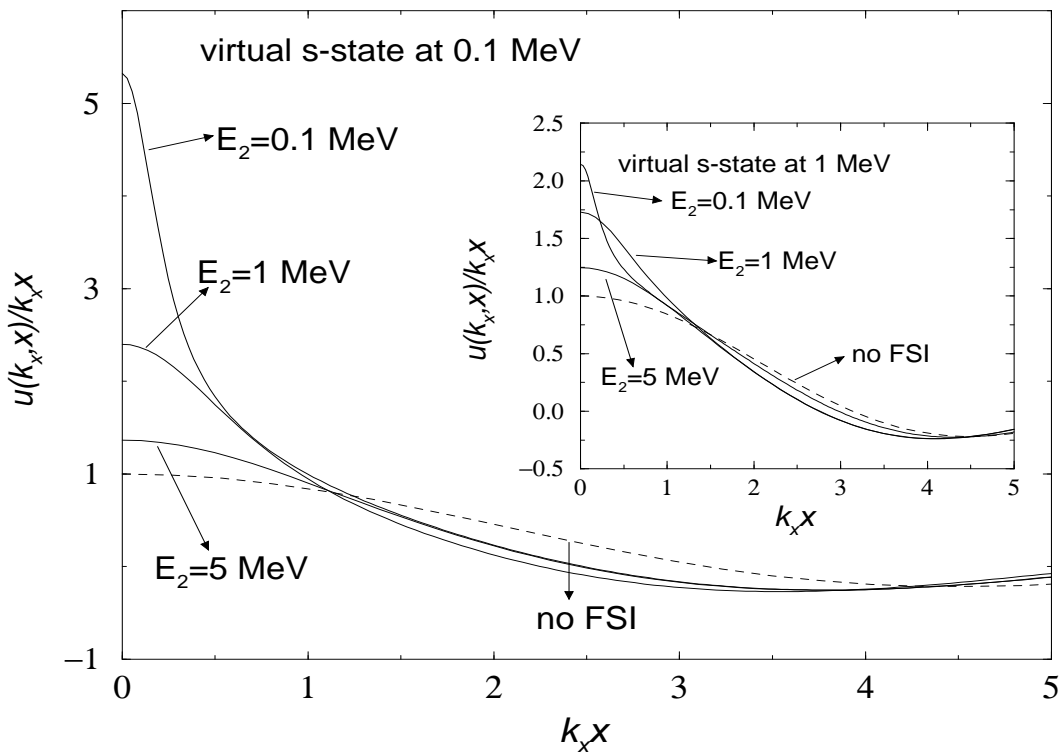}
\vspace{3.5cm}
\caption{\protect\small
Continuum radial $s$-state wave functions in
Eq.(\protect\ref{2bexp}) when the neutron-$^9$Li system has a virtual
$s$-state at 0.1 MeV. Three different two-body scattering energies
($E_2=\hbar^2 k_x^2 / 2m$) are considered, 0.1 MeV, 1 MeV and 5 MeV.
The dashed line is the same wave function when the neutron--$^9$Li
interaction is neglected. The inset show the same wave function when
the neutron-$^9$Li subsystem has a virtual $s$-state at 1 MeV.
}
\label{2a}
\end{figure}

To analyze the effect of the inclusion of the final state interaction
we investigate the behaviour of the distorted wave function in
Eq.(\ref{2bexp}).  First we consider a neutron-core potential with a
virtual $s$-state at 100 keV, and investigate how the corresponding
continuum radial $s$-state wave functions behave for different
energies $E_2 (=k_x^2 \hbar^2 /2m)$ of the neutron-core system.  When
the final state interaction is neglected this radial s-state wave
function is simply the Bessel function $j_0(k_x x)$, which in Fig. \ref{2a}
(the dashed line) is compared to the distorted wave functions (the
solid lines) for different values of $E_2$=0.1 MeV, 1 MeV and 5
MeV. At large distances (not shown in the figure) the oscillations
continue and have the same amplitude.

The difference from the Bessel function is only significant at small
distance and this difference increases in general when $E_2$
approaches the energy of the virtual state. (Note, however, the
singularity at the origin for vanishing $E_2$.)  Therefore for small
values of $k_x$ the overlap in Eq.(\ref{tran}) is larger than in
Eq.(\ref{tm}) and since the normalization is the same in all cases,
the momentum distribution becomes narrower by inclusion of final state
interactions.

The inset in Fig. \ref{2a} shows the same plot as in the main part, but for
a neutron-core potential with a virtual $s$-state at 1 MeV. The
general conclusions are the same, the smaller $E_2$ the larger the
effect of the final state interaction. However, in this case the
difference with the Bessel function is clearly smaller (note the
different vertical scales), and therefore also the influence of the final state
interaction is smaller. We can then conclude that the inclusion of
final state interaction in $s$-waves makes the momentum distribution
narrower, and the smaller the energy of the virtual $s$-state the
larger the effect of the final state interaction.

\begin{figure}[t]
\epsfxsize=12cm
\epsfysize=7cm
\epsfbox[650 200 1200 550]{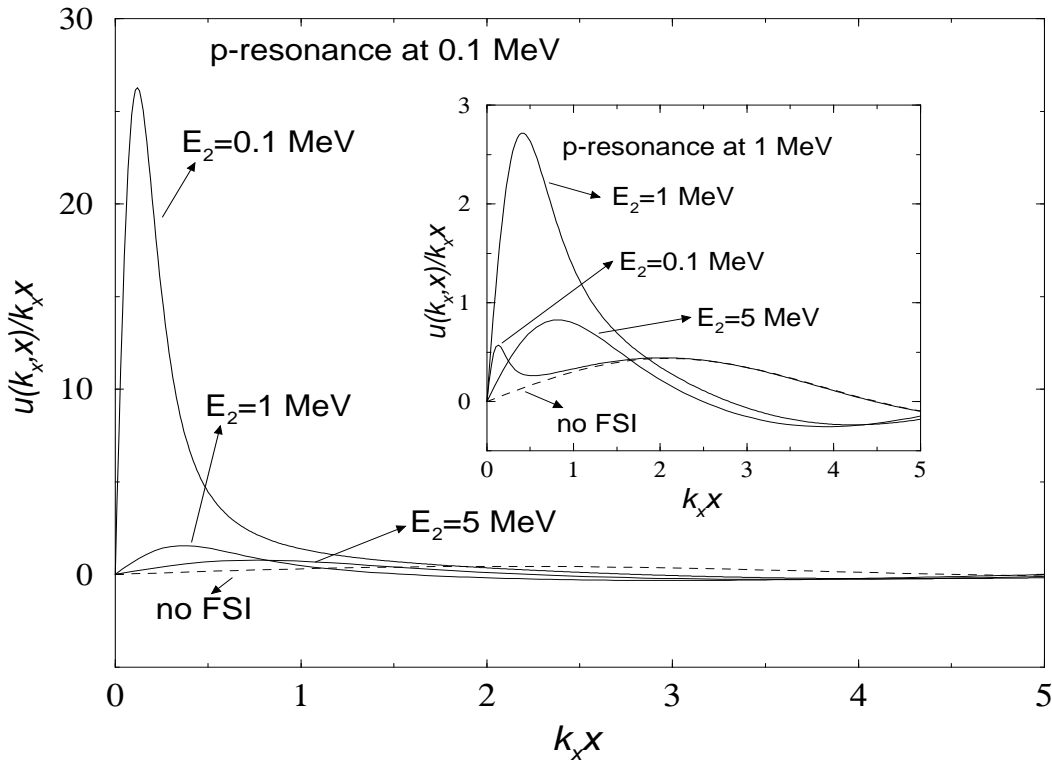}
\vspace{3.5cm}
\caption{\protect\small
The same as Fig. \protect\ref{2a} for p-resonances instead of virtual
s-states.
}
\label{2b}
\end{figure}

In Fig. \ref{2b} we show the same quantities as in Fig. \ref{2a}, but now for
$p$-waves. In the main part of the figure we consider a neutron-core
potential with a $p$-resonance at 0.1 MeV. Again the dashed line is the
radial part of the $p$-wave without final state interaction (equal to
the Bessel function $j_1(k_x x)$). When we solve the Schr\"{o}dinger
equation for a $p$-wave and when $E_2 (=k_x^2 \hbar^2 /2m)$ is far from
the resonance energy (we have taken $E_2=5$ MeV) the difference with
the Bessel function is very small, and the effect of the final state
interaction is almost negligible. When $E_2$ approaches the energy of
the resonance the $p$-wave becomes more and more localized, in such a
way that it gets a maximum for $E_2$ equal to the energy of the
resonance (0.1 MeV in this case). As a consequence, the overlap in
Eq.(\ref{tm}) is maximum for $k_x=\sqrt{2 m E_{\mbox{\scriptsize
res}}/\hbar^2}$, and the momentum distributions are enhanced around
that value of $k_x$.

The inset of Fig. \ref{2b} shows the same as in the main part for
$E_{\mbox{\scriptsize res}}$ = 1 MeV. The behaviour of the $p$-wave is
identical to the previous one. The wave is very much localized and
gets its maximum value for $E_2=E_{\mbox{\scriptsize res}}$. However,
the maximum is now clearly smaller than in the previous case of
$E_{\mbox{\scriptsize res}}$ = 0.1 MeV, and therefore also the effect
on the momentum distributions is less pronounced.  Therefore, the
inclusion of the final state interaction in the $p$-wave will produce
an increase in the momentum distributions around $k_x=\sqrt{2 m
E_{\mbox{\scriptsize res}}/\hbar^2}$. This increase is especially
important when the neutron-core potential has a low-lying
$p$-resonance, and again, due to the normalization, makes the momentum
distributions narrower.

Summarizing, the inclusion of final state interaction highly modifies
the momentum distributions for $s$- and $p$-waves. This change is
clearly more significant for low-lying virtual $s$-states and
low-lying $p$-resonances. For s-waves the distributions are enhanced
at the origin ($k_x=0$), and for p-waves the enhancement is produced
at $k_x=\sqrt{2 m E_{\mbox{\scriptsize res}}/\hbar^2}$. In both cases
the momentum distributions become narrower by inclusion of the final
state interaction.

Before concluding this subsection we must emphasize that the momentum
distributions discussed so far were expressed in terms of $k_x$,
i.e. the relative momentum between the two remaining particles.  From
Eqs.(\ref{mcm}) - (\ref{ab2}) we see that the momentum of the neutron
relative to the center of mass of the projectile is very close to
$\bbox{k}_x$ (the $b$ coefficient is small due to the large mass of
the core). Therefore the above conclusions about the effect of the
final state interaction roughly remain unchanged for the neutron
momentum distributions relative to the center of mass of the
projectile.  

For the core momentum distribution the $b_i$ coefficient in
Eq.(\ref{mcm}) is large, and the effect of the final state interaction
is much smaller when we refer momentum distributions of the core to
the center of mass of the projectile. The reason is simply that the
center of mass of the core is fairly close to the center of mass of
the total three-body system.

\subsection{Asymmetries}

Experimental data suggest that the neutron unbound $^{10}$Li system
has a $p$-resonance around 0.5 MeV \cite{you94} and a low-lying
($\leq$ 0.2 MeV) virtual $s$-state or $p$-resonance
\cite{zin95,you94,kryg93,boh93}.  In all the computations reported in
this work the neutron-core potential is fitted to produce a
$p$-resonance at 0.5 MeV in $^{10}$Li. For the lowest-lying state we
investigate several possibilities in order to analyze the dependence
of the polarization observables on the properties of the lowest-lying
state.

\begin{figure}[t]
\epsfxsize=12cm
\epsfysize=7cm
\epsfbox[550 200 1100 550]{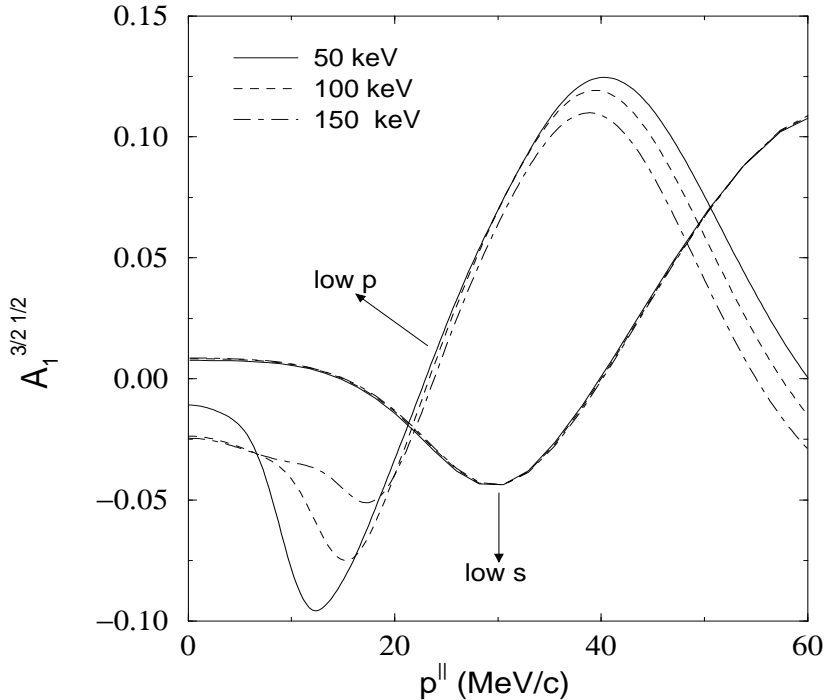}
\vspace{3.5cm}
\caption{\protect\small
Asymmetry $A_1^{3/2 \ 1/2}(p^\parallel)$ for neutron
momentum distributions in a neutron removal $^{11}$Li fragmentation
reaction. A low-lying state is assumed in $^{10}$Li at 50 keV (solid
line), 100 keV (dashed line) or at 150 keV (dot-dashed line). The
cases of low-lying $p$-resonances and low-lying virtual $s$-states are
both shown (see figure).
}
\label{3a}
\end{figure}

In Fig. \ref{3a} we show the asymmetry $A_1^{3/2 \ 1/2}(p^\parallel)$ for
neutron momentum distributions after a neutron removal $^{11}$Li
fragmentation reaction.  A low-lying state (s- or p-wave) is present
in $^{10}$Li at 50 keV (solid line), 100 keV (dashed line) or at 150
keV (dot-dashed line).  For low-lying $p$-resonances we observe for
all the three energies two peaks in the asymmetry. The first peak at
low momentum corresponds to the lowest-lying $p$-resonance. The second
peak is in all the three cases generated by the $p$-resonance at 0.5
MeV. The connection between the position of the different minima and
maxima of the asymmetry and the energy of the $p$-resonance may be
seen from the relations $\bbox{k}_x=\bbox{p}_r\sqrt{m/\mu}$ and $k_x =
\sqrt{2 m E_{\mbox{\scriptsize res}}/\hbar^2}$, where $p_r$ is the
relative neutron-core momentum, $E_{\mbox{\scriptsize res}}$ is the
energy of the resonance and $\mu$ is the reduced mass of the
neutron-core subsystem. The different p-resonance energies in the
figure correspond to values $p_r$ of 10 MeV/c, 14 MeV/c, and 17
MeV/c, respectively and an energy of 0.5 MeV corresponds to $p_r
\approx$ 30 MeV/c. 

As discussed in section 2, the momentum distributions are enhanced
around the resonance value of $k_x$, that is very close to the
momentum of the neutron relative to the center of mass of the
three-body projectile.  This increase in the momentum distributions
produces the oscillations in the asymmetry. As a consequence, when the
energy of the resonance increases the peak in the asymmetry is
displaced towards higher momenta.  Since we plot the asymmetries as
functions of the longitudinal component of the neutron momentum
relative to the center of mass of the projectile (that is not exactly
$p_r$, but very close), the maxima and minima of the asymmetry do not
exactly coincide with these numbers.  However, the connection to the
energy of the resonances in the neutron-core subsystem is very clear.

Inspecting the asymmetry in Fig. \ref{3a} where a low-lying virtual $s$-state
is present in $^{10}$Li we observe that the first peak completely
disappears.  This fact was expected, since in section 2 we showed that
the asymmetry Eq.(\ref{as1}) vanishes when only $s$-waves are involved
in the wave functions. In other words, the asymmetry $A_1^{3/2 \
1/2}(p^\parallel)$ does not feel the presence of virtual $s$-states in
the neutron-core subsystem, and it is only sensitive to the structure
of the non zero angular momnetum resonances. Since only s-waves are
changed for the three different cases considered, the asymmetries are
essentially indistinguishible.

\begin{figure}[t]
\epsfxsize=12cm
\epsfysize=7cm
\epsfbox[550 200 1100 550]{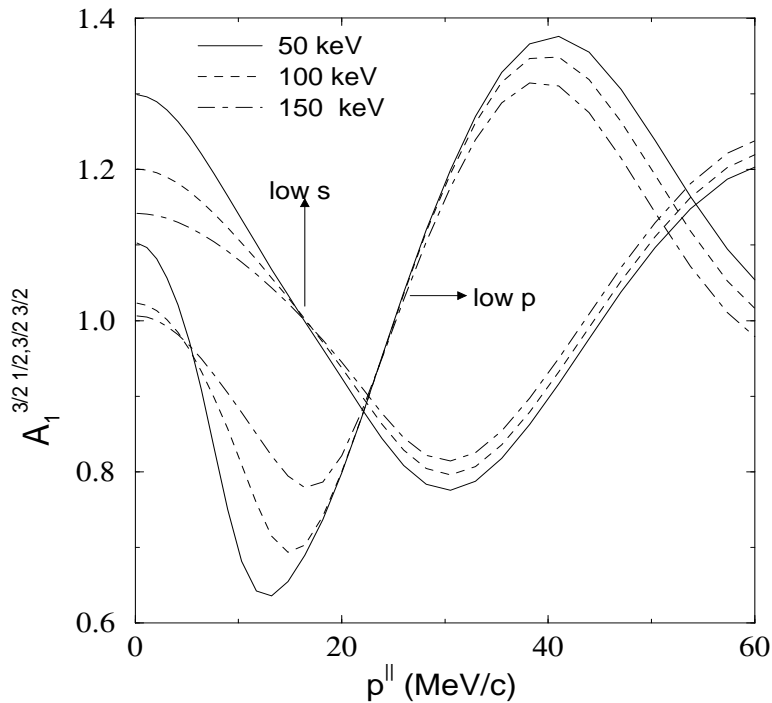}
\vspace{3.5cm}
\caption{\protect\small
The same as in Fig. \protect\ref{3a} for the asymmetry 
$A_1^{3/2 \ 1/2; 3/2 \ 3/2}(p^\parallel)$.
}
\label{4a}
\end{figure}

Let us now consider the case where the spin projection of one of the
particles in the final state is measured. In particular we assume that
the spin projection of the core is measured to be 3/2, and construct
the asymmetry $A_1^{3/2 \ 1/2;3/2 \ 3/2}(p^\parallel)$ for neutron
momentum distributions in a neutron removal reaction. The result is
shown in Fig. \ref{4a} for the cases corresponding to Fig. \ref{3a}.  Again, for
low-lying $p$-resonances there are two peaks in the asymmetry, one of
them connected to the energy of the low-lying resonance and the other
peak is related to the resonance at 0.5 MeV. 

For low-lying virtual $s$-states the first peak disappears, and only
the peak associated with the $p$-resonance at 0.5 MeV remains.  Now
the $s$-waves also contribute to the asymmetry, but the contribution
is localized in the region close to the origin where the curves now
are different for the three energies of the low-lying virtual
$s$-state.  Therefore the asymmetries shown in Fig. \ref{4a} display the
same features as those of Fig. \ref{3a}.  The main difference is that the
amplitude of the oscillations of the asymmetries clearly is larger in
the asymmetries shown in Fig. \ref{4a}.

\begin{figure}[t]
\epsfxsize=12cm
\epsfysize=7cm
\epsfbox[550 200 1100 550]{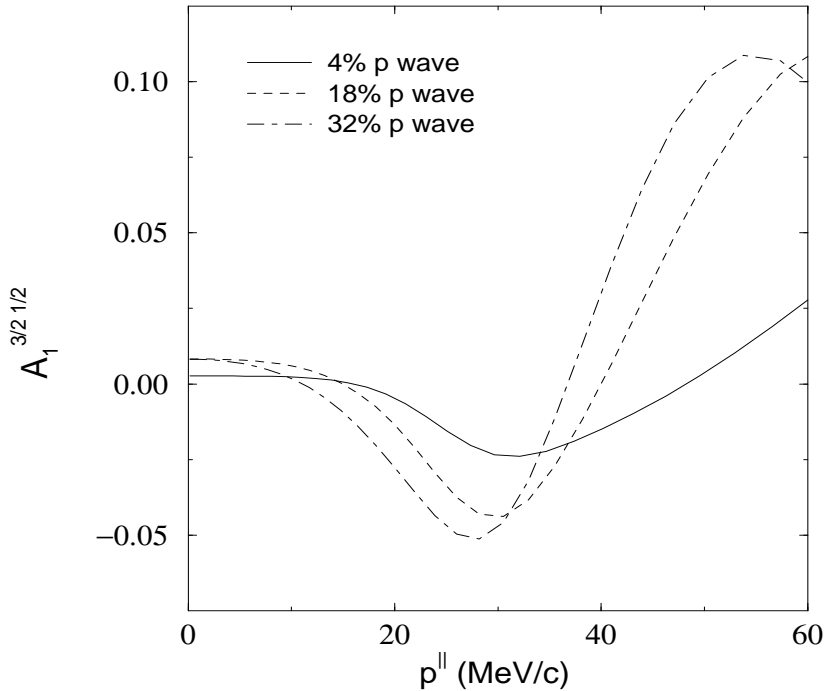}
\vspace{3.5cm}
\caption{\protect\small
Asymmetry $A_1^{3/2 \ 1/2}(p^\parallel)$ for neutron
momentum distributions in a neutron removal $^{11}$Li fragmentation
reaction. A low-lying virtual $s$-state and a $p$-resonance are assumed in
$^{10}$Li at 100 keV and 0.5 MeV, respectively. The $p$-wave content in
$^{10}$Li is considered to be 4\% (solid line), 18\% (dashed line), and
32\% (dot-dashed line). 
}
\label{5a}
\end{figure}

We have now concluded that the asymmetries are sensitive essentially
only to the non-zero angular momentum partial waves in the initial and
final state wave function.  Thus we also expect sensitivity to the
content of the different waves in the two-body subsystem. In
particular, for $^{11}$Li the asymmetries should depend on the $p$-wave
content in $^{10}$Li subsystem (the neutron-neutron subsystem contains
in any case almost exclusively $s$-waves). In Fig. \ref{5a} we show the same
asymmetry, $A_1^{3/2 \ 1/2}(p^\parallel)$, as in Fig. \ref{3a}, but now
assuming a virtual $s$-state at 100 keV and a $p$-resonance at 0.5 MeV
in $^{10}$Li. By use of the spin-orbit parameter $V^{l=1}_{so}$ in the
neutron-core potential in Eq.(\ref{potl}) we modify the $p$-wave
content in the neutron-$^9$Li subsystem.

The solid line, dashed line, and dot-dashed line are the resulting
calculations with 4\%, 18\%, and 32\% of $p$-wave content in the
$^{10}$Li wave function. In the three cases we observe the peak in the
asymmetry around 30 MeV/c corresponding to the $p$-resonance at 0.5
MeV. The second peak in the asymmetry corresponds to the next
$p$-resonance, closer to the first peak for a larger $p$-wave content,
since the increase in the $p$-wave content is obtained by decreasing
the energies of the $p$-resonances. 

We observe that the lower the $p$-wave content the smaller the
amplitudes of the oscillations of the asymmetry, see the solid line
corresponding to the $p$-wave content of 4\%. This is related to the
fact that the asymmetry must vanish when the $p$-waves are totally
absent, However, it should be mentioned that the denominator of the
asymmetry in Eq.(\ref{as1}) is the unpolarized momentum distribution, that
is known to be narrower for lower $p$-wave contents \cite{gar96}. It
is then possible to find oscillations with similar amplitude for
different values of the $p$-wave content.

\vspace{-7mm}
\begin{figure}[t]
\epsfxsize=12cm
\epsfysize=7cm
\epsfbox[550 200 1100 550]{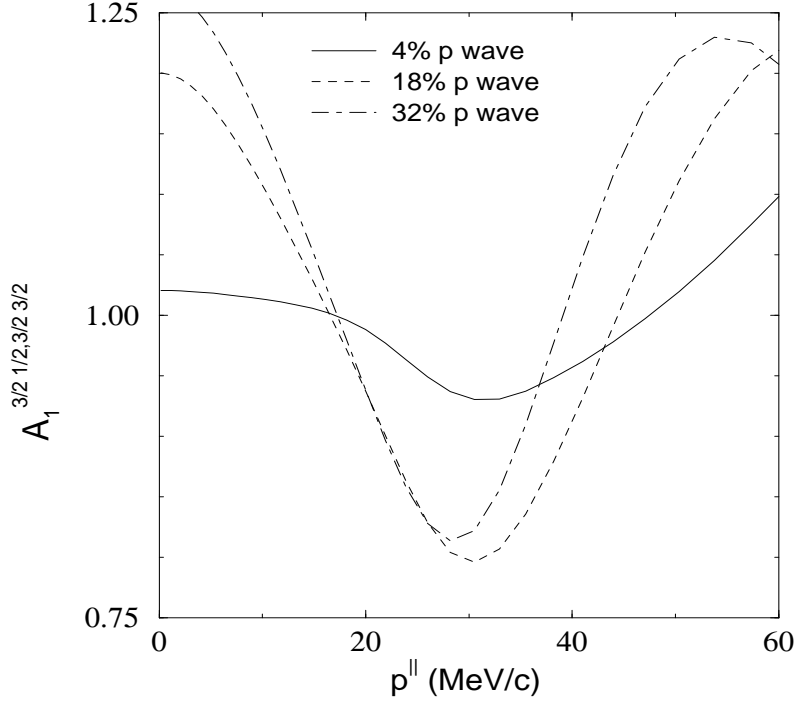}
\vspace{3.1cm}
\caption{\protect\small
The same as in Fig. \protect\ref{5a} for the asymmetry
$A_1^{3/2 \ 1/2; 3/2 \ 3/2}(p^\parallel)$.
}
\label{6a}
\end{figure}

In Fig. \ref{6a} we show the same asymmetry, $A_1^{3/2 \ 1/2; 3/2 \
3/2}(p^\parallel)$, as in Fig. \ref{5a}, i.e. one-dimensional neutron
momentum distributions in a neutron removal reaction. As in Fig. \ref{4a} the
spin projection of the core is assumed to be measured to have a value
of 3/2 in the final state. Again, in this case the asymmetry shows
essentially the same features as in Fig. \ref{5a}, but with the advantage of
larger amplitudes. The values of the asymmetries can vary now between
0.8 and more than 1.3.

\vspace{-6mm}
\begin{figure}[t]
\epsfxsize=12cm
\epsfysize=7cm
\epsfbox[550 200 1100 550]{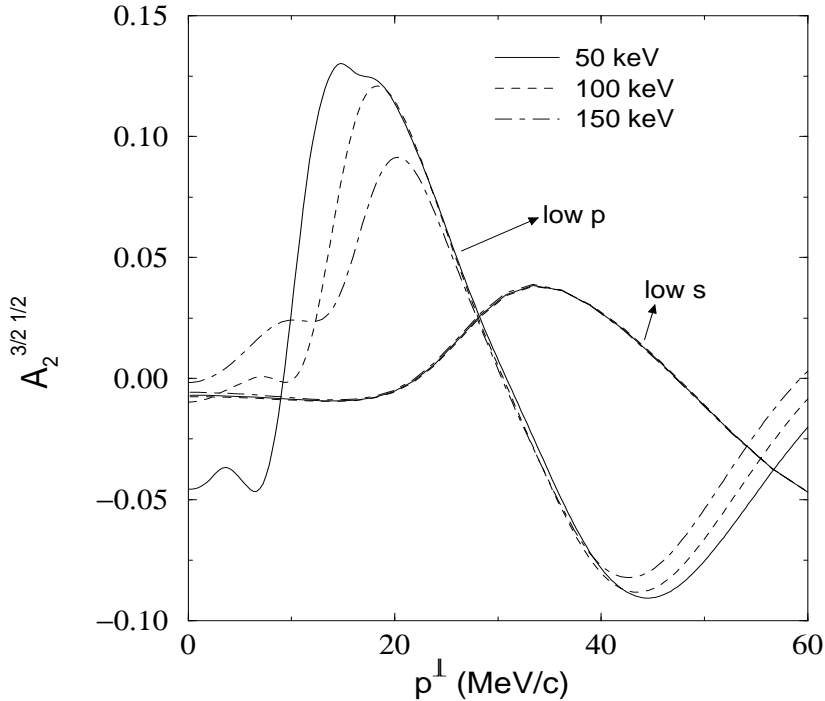}
\vspace{2.9cm}
\caption{\protect\small
The same as in Fig. \protect\ref{3a} for the asymmetry 
$A_2^{3/2 \ 1/2}(p^\bot)$.
}
\label{3b}
\end{figure}

\begin{figure}[t]
\epsfxsize=12cm
\epsfysize=7cm
\epsfbox[550 200 1100 550]{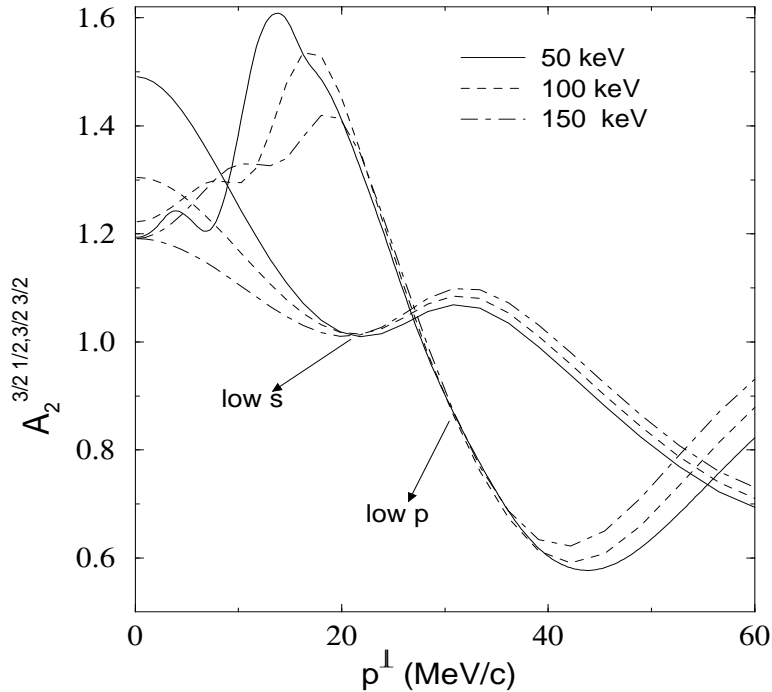}
\vspace{3.5cm}
\caption{\protect\small
 The same as in Fig. \protect\ref{4a} for the asymmetry
$A_2^{3/2 \ 1/2; 3/2 \ 3/2}(p^\bot)$.
}
\label{4b}
\end{figure}

\begin{figure}[t]
\epsfxsize=12cm
\epsfysize=7cm
\epsfbox[550 200 1100 550]{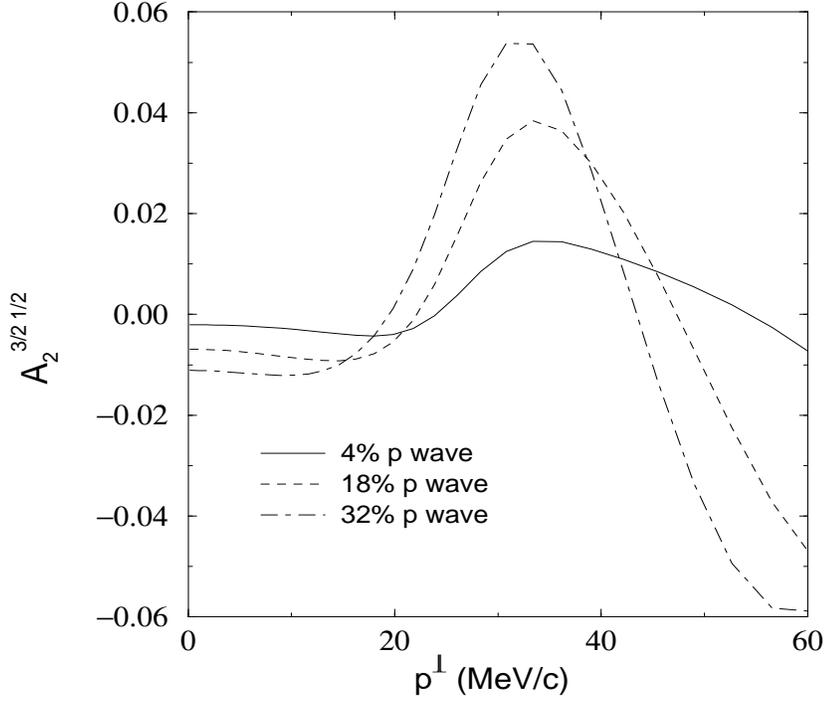}
\vspace{3.1cm}
\caption{\protect\small
The same as in Fig. \protect\ref{5a} for the asymmetry
$A_2^{3/2 \ 1/2}(p^\bot)$.
}
\label{5b}
\end{figure}

\begin{figure}[t]
\epsfxsize=12cm
\epsfysize=7cm
\epsfbox[550 200 1100 550]{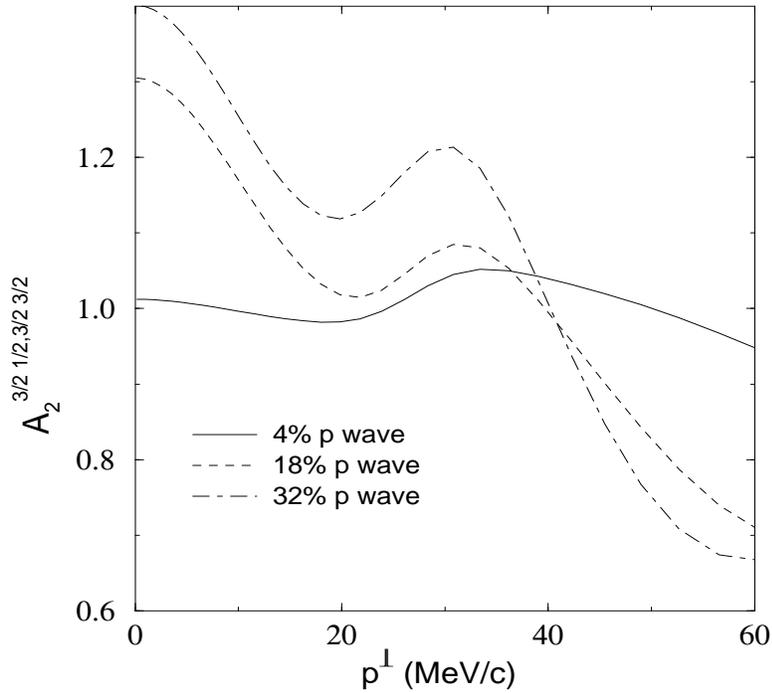}
\vspace{3.1cm}
\caption{\protect\small
The same as in Fig. \protect\ref{6a} for the asymmetry
$A_2^{3/2 \ 1/2; 3/2 \ 3/2}(p^\bot)$.
}
\label{6b}
\end{figure}

All the asymmetries we have discussed so far are obtained from
one-dimensional neutron momentum distributions.  The same conclusions
can be obtained from the corresponding two-dimensional neutron momentum
distributions shown for completeness in Figs. \ref{3b}, \ref{4b}, 
\ref{5b}, \ref{6b}.  First we
note that when only the beam is polarized the asymmetries are not
sensitive to the structure of the virtual $s$-states in the two-body
subsystems. When only the energy of the $s$-states is modified in the
$^{10}$Li wave function the asymmetries do not change. Second, the
minima and maxima of the asymmetries provide information about the
$p$-resonance structure in the $^{10}$Li two-body subsystem. Third, the
asymmetries are clearly dependent on the $p$-wave content of the
$^{10}$Li wave function. Fourth, the asymmetries obtained assuming that
the spin projection of the core has been measured in the final state
(3/2 in our case) show a clearly larger effect than those obtained
assuming that only the beam is polarized.

In principle the same asymmetries could also be obtained from core
momentum distributions and from neutron momentum distributions after
core break-up. However these two cases are not especially interesting.
The former because core momentum distributions are very little affected
by the final state interaction, and then the asymmetries do not show
very much structure. The latter because after core break-up the two
neutrons survive in the final state, and the three-body wave function
written in the Jacobi set where the $\bbox{x}$ coordinate is drawn
between the two neutrons contains almost exclusively s-waves, to which
the asymmetries are not sensitive.

\section{Summary and conclusions}

We have formulated a detailed model to investigate momentum
distributions of particles emerging after nuclear break-up reactions
of fast polarized three-body halo nuclei. The three main ingredients
are:

(i) The three-body description of the initial nuclear wave function.
This assumes that the particles or clusters are inert throughout the
process and only distances larger than the cluster radii can be well
described. Such models have been successfully applied to various light
nuclei.

(ii) The sudden approximation where the target nucleus instantaneously
removes one of the particles from the three-body projectile without
disturbing the other two. This approximation is very well suited for
short range interactions and fast processes where the beam velocity is
much larger than the velocity of the intrinsic motion. The radioactive
beams are available with high energies exceedingly well fulfilling
this criterium.

(iii) The final state relative wave function for the two remaining
particles must include effects of their mutual interaction. The
resulting distorted wave function may differ substantially from the
plane wave solution when no two-body interaction is present. These
effects are crucial when virtual s-states or higher angular momentum
resonances occur at energies below about 1 MeV in the final state
two-body system. 

\vspace{-5mm}

It is rather difficult to construct a theoretical system where
assumption (i) is valid without the necessity of using (iii). The
reason is that the three-body structure only is appropriate when the
the particles are fairly loosely bound, since they otherwise would
overlap and destroy the cluster property. As a consequence the
two-body subsystems must have fairly weak effective attractions yet
they must be sufficiently strong to create a three-body bound state
when working together. The result is low-lying continuum two-body
structure.

The same two-body interactions are therefore responsible for both the
initial three-body bound state and for the distorted final state wave
function. A consistent treatment of initial and final stages of the
process is therefore essential and our model meets this crucial
requirement. In other words the two-body continuum structure is
decisive for both the three-body halo structure and the momentum
distributions after fragmentation.

We first describe the method to compute the momentum distributions in
nuclear break-up processes. The overlap between initial and final state
wave function is the essential quantity. The initial bound state
three-body halo wave function is obtained by solving the Faddeev
equations by use of an adiabatic hyperspherical expansion technique.
This method has been tested previously and shown to be very efficient
for ground states or low-lying excited states with small two-body
relative orbital angular momenta. The final state distorted two-body
continuum wave function is expanded in partial waves and the related
radial wave functions are calculated numerically. They are normalized
at large distances to coincide with the plane wave solutions. The
resulting overlap matrix with different angular momentum projections
is then available.

The cross sections or the observable momentum distributions are then
essentially only the absolute square of the overlap integrated over
unobserved quantities.  These expressions contain numerically computed
quantities. They appear to be rather complicated especially due to the
use of the coordinate system where the center of mass of the
three-body projectile is at rest and the dedicated goal of computing
momentum distributions for polarized projectiles.

First we specify the two-body interactions appropriate for the
$^{11}$Li system considered as two neutrons surrounding the
$^{9}$Li-core. The neutron-neutron interaction is parametrized to
reproduce low-energy scattering properties and then maintained
throughout the present investigation. The neutron-core interaction is
adjusted to reproduce the measured $^{11}$Li binding energy and root
mean square radius. The remaining free parameters are used to place
virtual s-states and p-resonances at various energies close to the
threshold resulting in different $^{11}$Li-structures and different
momentum distributions.

The final state two-body s-wave functions all have maxima at the
origin. The highest value occurs when the scattering energy equals to
the energy of the virtual s-state. The state is therefore peaked around
zero and this localization is strongest around the energy of the
virtual state.  The p-state two-body wave function is peaked at an
energy corresponding to the p-resonance energy. Without resonances and
virtual states the wave functions are simply Bessel functions with the
related oscillations and without these pronounced peak structures.

Neutron momentum distributions after neutron removal are computed for
polarized $^{11}$Li and when the polarization of the final two-body
state is measured.  Several polarization observables are constructed
and expressed in terms of quantities, called asymmetries, deviating
from zero only when the momentum distributions depend on the
polarization. Several continuum structures of $^{10}$Li are considered,
but in all cases we maintain the binding energy and root mean square
radius of $^{11}$Li.

The inclusion of the final state interaction is an essential ingredient
in the computation, determining the behaviour of the asymmetries.  In
fact, when they are not included the structure in the polarization
observables completely disappears.  The core momentum distributions are
little affected by the final state interaction, and therefore the
asymmetries obtained from them exhibit very little structure.

From the asymmetries constructed from neutron momentum distributions we
have observed the following features: If only the beam is polarized the
asymmetries vanish when only s-waves contribute. This makes the
asymmetries obtained from a core break-up reaction from $^{11}$Li not
relevant, because almost exclusively s-waves are involved in this
fragmentation process.  Much more interesting are the asymmetries
obtained from neutron momentum distributions after neutron removal
reactions. The rest of the waves plays an important role, and the
asymmetries become a perfect tool to investigate the $\ell >
0$-resonance structure of the neutron-core subsystem. These observables
then contain information about the continuum structure of the two-body
subsystems.  In fact low-lying resonances (p-resonances in the
$^{11}$Li case) in the neutron-core system show up as an extremum at
the momentum value corresponding to the resonance energy.  An additonal
measurement of the two-body final state polarization enhances the
amplitudes of the asymmetries, but otherwise leaves essentially
unchanged the structure of the asymmetry. 

\appendix
\section{Relative Momentum Distributions}
In this appendix we give the analytic expressions for the relative
momentum distributions of the two particles in the final state. We
consider three different cases:

(i) 100\% polarized beam and spin projection of one of the particles in the
final state measured:

\begin{small}
\begin{eqnarray}
\frac{d^3 \sigma}{dk_y dk_x^\bot dk_x^\parallel} & \propto &
k_y^2 k_x^\bot \frac{2}{\pi^2} \sum_{I L_x} (-1)^{J-M} \hat{I}^2
\left(
    \begin{array}{ccc}
       J & J & I \\
       M & -M & 0
    \end{array}
    \right) P_{L_x}(\cos{\theta_{k_x}})
           \nonumber \\ & &  \hspace{-3cm}
\sum_{j_x \ell_x s_x j_x^\prime \ell_x^\prime s_x^\prime}
\sum_{L S L^\prime S^\prime} \sum_{\ell_y j_y}
(-1)^{\ell_x+J-s_x^\prime+2s_y-j_y+j_x-s_2-s_1}
\hat{j_x}^2 \hat{j^\prime_x}^2 \hat{j_y}^2 \hat{J}^2 \hat{\ell_x}
\hat{\ell^\prime_x} \hat{L} \hat{L^\prime} \hat{S} \hat{S^\prime}
\hat{s_x} \hat{s^\prime_x} \hat{L_x}^2 \hat{s_1}^2
    \nonumber \\ & & \hspace{-3cm}
I_{\ell_x s_x j_x}^{\ell_y L S}(\kappa, \alpha_\kappa)
I_{\ell_x^\prime s_x^\prime j_x^\prime}^{\ell_y L^\prime S^\prime}
(\kappa, \alpha_\kappa)
\left\{
    \begin{array}{ccc}
       J & j_x & j_y \\
       L & \ell_x & \ell_y \\
       S & s_x & s_y
    \end{array}
    \right\}
\left\{
    \begin{array}{ccc}
       J & j_x^\prime & j_y \\
       L^\prime & \ell_x^\prime & \ell_y \\
       S^\prime & s_x^\prime & s_y
    \end{array}
    \right\}
\left\{
    \begin{array}{ccc}
       j_x & j_x^\prime & I \\
       J & J & j_y
    \end{array}
    \right\}
\left(
    \begin{array}{ccc}
       \ell_x & \ell_x^\prime & L_x \\
       0 & 0 & 0
    \end{array}
    \right)
              \nonumber \\ & & \hspace{-3cm}
\sum_{I^\prime} (-1)^{I^\prime+s_1-\sigma_1} \hat{I^\prime}^2
\left(
    \begin{array}{ccc}
       s_1 & s_1 & I^\prime \\
       \sigma_1 & -\sigma_1 & 0
    \end{array}
    \right)
\left(
    \begin{array}{ccc}
       I^\prime & I & L_x \\
       0 & 0 & 0
    \end{array}
    \right)
\left\{
    \begin{array}{ccc}
       s_1 & I^\prime & s_1 \\
       s_x^\prime & s_2 & s_x
    \end{array}
    \right\}
\left\{
    \begin{array}{ccc}
       s_x & s_x^\prime & I^\prime \\
       j_x & j_x^\prime & I \\
       \ell_x & \ell_x^\prime & L_x
    \end{array}
    \right\}
\label{rel1}
\end{eqnarray}
\end{small}

(ii) 100\% polarized beam and no spins projections measured in the 
final state:

  This amounts to multiply the previous expression by $1/(2s_1+1)$ and
sum over $\sigma_1$. The main consequence of the summation is that the
index $I^\prime$ is zero, and then $L_x=I$. Then we get the expression

\begin{small}
\begin{eqnarray}
\frac{d^3 \sigma}{dk_y dk_x^\bot dk_x^\parallel} & \propto &
k_y^2 k_x^\bot \frac{2}{\pi^2} \sum_I (-1)^{J-M} \hat{I}^2
\left(
    \begin{array}{ccc}
       J & J & I \\
       M & -M & 0
    \end{array}
    \right) P_I(\cos{\theta_{k_x}})
\label{rel2} \\ & & \hspace{-3cm}
\sum_{j_x \ell_x s_x j_x^\prime \ell_x^\prime }
\sum_{L S L^\prime S^\prime} \sum_{\ell_y j_y}
(-1)^{2s_y+J-s_x+j_x-j_x^\prime-j_y}
\hat{j_x}^2 \hat{j^\prime_x}^2 \hat{j_y}^2 \hat{J}^2 \hat{\ell_x}
\hat{\ell^\prime_x} \hat{L} \hat{L^\prime} \hat{S} \hat{S^\prime}
I_{\ell_x s_x j_x}^{\ell_y L S}(\kappa, \alpha_\kappa)
    \nonumber \\ & & \hspace{-3cm}
I_{\ell_x^\prime s_x j_x^\prime}^{\ell_y L^\prime S^\prime}
(\kappa, \alpha_\kappa)
\left\{
    \begin{array}{ccc}
       J & j_x & j_y \\
       L & \ell_x & \ell_y \\
       S & s_x & s_y
    \end{array}
    \right\}
\left\{
    \begin{array}{ccc}
       J & j_x^\prime & j_y \\
       L^\prime & \ell_x^\prime & \ell_y \\
       S^\prime & s_x & s_y
    \end{array}
    \right\}
\left(
    \begin{array}{ccc}
       \ell_x & \ell_x^\prime & I \\
       0 & 0 & 0
    \end{array}
    \right)
\left\{
    \begin{array}{ccc}
       j_x & j_x^\prime & I \\
       J & J & j_y
    \end{array}
    \right\}
\left\{
    \begin{array}{ccc}
       j_x & j_x^\prime & I \\
       \ell_x^\prime & \ell_x & s_x
    \end{array}
    \right\} \nonumber
\end{eqnarray}
\end{small}

(iii) Unpolarized beam and no spins projections measured in the final state:

  Multiplying by $1/(2J+1)$ and summing over $M$ we get the totally
unpolarized momentum distribution. From the summation we obtain that
$I=0$, and the analytic expression is 

\begin{small}
\begin{eqnarray}
\frac{d^3 \sigma}{dk_y dk_x^\bot dk_x^\parallel} & \propto &
k_y^2 k_x^\bot \frac{2}{\pi^2} 
\sum_{j_x \ell_x s_x }
\sum_{L S L^\prime S^\prime} \sum_{\ell_y j_y}
\hat{j_x}^2 \hat{j_y}^2 \hat{L} \hat{L^\prime} \hat{S} \hat{S^\prime}
I_{\ell_x s_x j_x}^{\ell_y L S}(\kappa, \alpha_\kappa)
I_{\ell_x s_x j_x}^{\ell_y L^\prime S^\prime}
(\kappa, \alpha_\kappa)
    \nonumber \\ & &
\left\{
    \begin{array}{ccc}
       J & j_x & j_y \\
       L & \ell_x & \ell_y \\
       S & s_x & s_y
    \end{array}
    \right\}
\left\{
    \begin{array}{ccc}
       J & j_x & j_y \\
       L^\prime & \ell_x & \ell_y \\
       S^\prime & s_x & s_y
    \end{array}
    \right\}
\label{rel3}
\end{eqnarray}
\end{small}

\section{Momentum Distributions Relative to the Center of Mass of the
Projectile}
In this appendix we give the analytic expressions for the 
momentum distributions of the two particles in the final state relative
to the center of mass of the incident three-body system. The same
three cases as in the previous appendix are considered: 

(i) 100\% polarized beam and spin projection of one of the particles in the
final state measured:

\begin{small}
\begin{eqnarray}
\frac{d^4\sigma}{dk_y d\theta^\prime_{k_y} dp^\bot dp^\parallel}
   & \propto & \sin{\theta^\prime_{k_y}} k_y^2 p^\bot \frac{1}{a_i^3 \pi^2}
\sum_I (-1)^{J-M} \hat{I}^2 
\left(
    \begin{array}{ccc}
       J & J & I \\
       M & -M & 0
    \end{array}
    \right) 
\sum_{\ell_x s_x j_x L S}
         \nonumber \\ & & \hspace{-2.5cm}
\sum_{\ell_x^\prime s_x^\prime j_x^\prime L^\prime S^\prime}
\sum_{\ell_y j_y \ell_y^\prime j_y^\prime}
(-1)^{s_y-\ell_x+2j_x^\prime+j_y^\prime-s_2-s_1+s_x^\prime}
\hat{j_x}^2 \hat{j^\prime_x}^2 \hat{j_y}^2 \hat{j_y^\prime}^2 
\hat{J}^2 \hat{\ell_x} \hat{\ell^\prime_x} \hat{\ell_y}
\hat{\ell^\prime_y} \hat{L} \hat{L^\prime} \hat{S} \hat{S^\prime}
         \nonumber \\ & & \hspace{-2.5cm}
\hat{s_x} \hat{s^\prime_x} \hat{s_1}^2
I_{\ell_x s_x j_x}^{\ell_y L S}(\kappa, \alpha_\kappa)
I_{\ell_x^\prime s_x^\prime j_x^\prime}^{\ell_y^\prime L^\prime S^\prime}
(\kappa, \alpha_\kappa)
\left\{
    \begin{array}{ccc}
       J & j_x & j_y \\
       L & \ell_x & \ell_y \\
       S & s_x & s_y
    \end{array}
    \right\}
\left\{
    \begin{array}{ccc}
       J & j_x^\prime & j_y^\prime \\
       L^\prime & \ell_x^\prime & \ell_y^\prime \\
       S^\prime & s_x^\prime & s_y
    \end{array}
    \right\}
\sum_{L_x L_y} \hat{L_x}^2 \hat{L_y}^2
         \nonumber \\ & & \hspace{-2.5cm}
\left(
    \begin{array}{ccc}
       \ell_x & \ell_x^\prime & L_x \\
       0 & 0 & 0
    \end{array}
    \right) 
\left(
    \begin{array}{ccc}
       \ell_y & \ell_y^\prime & L_y \\
       0 & 0 & 0
    \end{array}
    \right) 
\left\{
    \begin{array}{ccc}
       j_y & j^\prime_y & L_y \\
       \ell^\prime_y & \ell_y & s_y
    \end{array}
    \right\} 
\sum_{T N_x} \hat{T}^2 (-1)^T P_T(\cos{\theta_p})
         \nonumber \\ & & \hspace{-2.5cm}
\left(
    \begin{array}{ccc}
       L_x & L_y & T \\
       N_x & -N_x & 0
    \end{array}
    \right) 
\sqrt{\frac{(L_x-N_x)! (L_y+N_x)!}{(L_x+N_x)! (L_y-N_x)!}}
P_{L_x}^{N_x}(\cos{\theta^\prime_{k_x}})
P_{L_y}^{-N_x}(\cos{\theta^\prime_{k_y}})
         \nonumber \\ & & \hspace{-2.5cm}
\sum_{I^\prime I^{''}} (-1)^{s_1-\sigma_1+I^\prime-I^{''}} 
\hat{I^\prime}^2 \hat{I^{''}}^2
\left(
    \begin{array}{ccc}
       s_1 & s_1 & I^\prime \\
       \sigma_1 & -\sigma_1 & 0
    \end{array}
    \right) 
\left\{
    \begin{array}{ccc}
       j_x & j_x^\prime & I^{''} \\
       j_y & j_y^\prime & L_y \\
       J & J & I
    \end{array}
    \right\} 
\left\{
    \begin{array}{ccc}
       s_x & s_x^\prime & I^\prime \\
       j_x & j_x^\prime & I^{''} \\
       \ell_x & \ell^\prime_x & L_x
    \end{array}
    \right\} 
         \nonumber \\ & & \hspace{-2.5cm}
\left\{
    \begin{array}{ccc}
       s_1 & I^\prime & s_1 \\
       s_x^\prime & s_2 & s_x
    \end{array}
    \right\} 
\left(
    \begin{array}{ccc}
       I & T & I^\prime \\
       0 & 0 & 0 
    \end{array}
    \right) 
\left\{
    \begin{array}{ccc}
       I & T & I^\prime \\
       L_x & I^{''} & L_y 
    \end{array}
    \right\} 
\label{cm1}
\end{eqnarray}
\end{small}

(ii) 100\% polarized beam and no spins projections measured in the 
final state:

\begin{small}
\begin{eqnarray}
\frac{d^4\sigma}{dk_y d\theta^\prime_{k_y} dp^\bot dp^\parallel}
   & \propto & \sin{\theta^\prime_{k_y}} k_y^2 p^\bot \frac{1}{a_i^3 \pi^2}
\sum_I (-1)^{J-M} \hat{I}^2 P_I(\cos{\theta_p})
\left(
    \begin{array}{ccc}
       J & J & I \\
       M & -M & 0
    \end{array}
    \right) 
         \nonumber \\ & & \hspace{-2.5cm}
\sum_{\ell_x s_x j_x L S}
\sum_{\ell_x^\prime j_x^\prime L^\prime S^\prime}
\sum_{\ell_y j_y \ell_y^\prime j_y^\prime}
(-1)^{s_y+j_x^\prime+j_y^\prime+s_x+I}
\hat{j_x}^2 \hat{j^\prime_x}^2 \hat{j_y}^2 \hat{j_y^\prime}^2 
\hat{J}^2 \hat{\ell_x} \hat{\ell^\prime_x} \hat{\ell_y}
\hat{\ell^\prime_y} \hat{L} \hat{L^\prime} \hat{S} \hat{S^\prime}
         \nonumber \\ & & \hspace{-2.5cm}
I_{\ell_x s_x j_x}^{\ell_y L S}(\kappa, \alpha_\kappa)
I_{\ell_x^\prime s_x j_x^\prime}^{\ell_y^\prime L^\prime S^\prime}
(\kappa, \alpha_\kappa)
\left\{
    \begin{array}{ccc}
       J & j_x & j_y \\
       L & \ell_x & \ell_y \\
       S & s_x & s_y
    \end{array}
    \right\}
\left\{
    \begin{array}{ccc}
       J & j_x^\prime & j_y^\prime \\
       L^\prime & \ell_x^\prime & \ell_y^\prime \\
       S^\prime & s_x & s_y
    \end{array}
    \right\}
\sum_{L_x L_y} (-1)^{L_x+L_y} \hat{L_x}^2 \hat{L_y}^2
         \nonumber \\ & & \hspace{-2.5cm}
\left(
    \begin{array}{ccc}
       \ell_x & \ell_x^\prime & L_x \\
       0 & 0 & 0
    \end{array}
    \right) 
\left(
    \begin{array}{ccc}
       \ell_y & \ell_y^\prime & L_y \\
       0 & 0 & 0
    \end{array}
    \right) 
\left\{
    \begin{array}{ccc}
       j_y & j^\prime_y & L_y \\
       \ell^\prime_y & \ell_y & s_y
    \end{array}
    \right\} 
\left\{
    \begin{array}{ccc}
       j_x & j_x^\prime & L_x \\
       j_y & j_y^\prime & L_y \\
       J & J & I
    \end{array}
    \right\} 
\left\{
    \begin{array}{ccc}
       j_x & j_x^\prime & L_x \\
       \ell_x^\prime & \ell_x & s_x
    \end{array}
    \right\} 
         \nonumber \\ & & \hspace{-2.5cm}
\sum_{N_x} 
\left(
    \begin{array}{ccc}
       L_x & L_y & I \\
       N_x & -N_x & 0
    \end{array}
    \right) 
\sqrt{\frac{(L_x-N_x)! (L_y+N_x)!}{(L_x+N_x)! (L_y-N_x)!}}
P_{L_x}^{N_x}(\cos{\theta^\prime_{k_x}})
P_{L_y}^{-N_x}(\cos{\theta^\prime_{k_y}})
\label{cm2}
\end{eqnarray}
\end{small}

(iii) Unpolarized beam and no spins projections measured in the final state:

\begin{small}
\begin{eqnarray}
\frac{d^4\sigma}{dk_y d\theta^\prime_{k_y} dp^\bot dp^\parallel}
   & \propto & \sin{\theta^\prime_{k_y}} k_y^2 p^\bot \frac{1}{a_i^3 \pi^2}
\sum_{\ell_x s_x j_x L S}
\sum_{\ell_x^\prime j_x^\prime L^\prime S^\prime}
\sum_{\ell_y j_y \ell_y^\prime j_y^\prime}
(-1)^{s_y-s_x+j_y^\prime+j_y+J}
         \nonumber \\ & & \hspace{-2.5cm}
\hat{j_x}^2 \hat{j^\prime_x}^2 \hat{j_y}^2 \hat{j_y^\prime}^2 
\hat{\ell_x} \hat{\ell^\prime_x} \hat{\ell_y}
\hat{\ell^\prime_y} \hat{L} \hat{L^\prime} \hat{S} \hat{S^\prime}
I_{\ell_x s_x j_x}^{\ell_y L S}(\kappa, \alpha_\kappa)
I_{\ell_x^\prime s_x j_x^\prime}^{\ell_y^\prime L^\prime S^\prime}
(\kappa, \alpha_\kappa)
\left\{
    \begin{array}{ccc}
       J & j_x & j_y \\
       L & \ell_x & \ell_y \\
       S & s_x & s_y
    \end{array}
    \right\}
         \nonumber \\ & & \hspace{-2.5cm}
\left\{
    \begin{array}{ccc}
       J & j_x^\prime & j_y^\prime \\
       L^\prime & \ell_x^\prime & \ell_y^\prime \\
       S^\prime & s_x & s_y
    \end{array}
    \right\}
\sum_{L_x} \hat{L_x}^2 
\left(
    \begin{array}{ccc}
       \ell_x & \ell_x^\prime & L_x \\
       0 & 0 & 0
    \end{array}
    \right) 
\left(
    \begin{array}{ccc}
       \ell_y & \ell_y^\prime & L_x \\
       0 & 0 & 0
    \end{array}
    \right) 
\left\{
    \begin{array}{ccc}
       j_y & j^\prime_y & L_x \\
       \ell^\prime_y & \ell_y & s_y
    \end{array}
    \right\} 
\left\{
    \begin{array}{ccc}
       j_x & j^\prime_x & L_x \\
       \ell^\prime_x & \ell_x & s_x
    \end{array}
    \right\} 
         \nonumber \\ & & \hspace{-2.5cm}
\left\{
    \begin{array}{ccc}
       j_x & j^\prime_x & L_x \\
       j^\prime_y & j_y & J
    \end{array}
    \right\} 
\sum_{N_x} (-1)^{N_x}
P_{L_x}^{N_x}(\cos{\theta^\prime_{k_x}})
P_{L_x}^{-N_x}(\cos{\theta^\prime_{k_y}})
\label{cm3}
\end{eqnarray}
\end{small}

All the three previous expressions are valid when $a<0$. For $a>0$ the
summation over $N_x$ should include an extra factor $(-1)^{N_x}$.

If we assume that one of the particles in the final state, say particle
1, has infinite mass, then the momentum distribution of particle 2
relative to the three-body center of mass should coincide with the
momentum distribution relative to particle 1. In other words,
Eqs.(\ref{rel1}), (\ref{rel2}), and (\ref{rel3}) should be recovered by
integrating over $\theta^\prime_{k_y}$ Eqs.(\ref{cm1}), (\ref{cm2}),
and (\ref{cm3}), respectively. In this case $b_i=0$, see
Eq.(\ref{ab2}), $\bbox{p} = a_i\bbox{k}_x$ and therefore
$\theta^\prime_{k_x}=0$ and $P_{L_x}^{N_x}(1)=\delta_{N_x,0}$.  Now
$\kappa$ and $\alpha_\kappa$ are independent of $\theta^\prime_{k_y}$
and the integration over
$\theta^\prime_{k_y}$ can easily be done analytically leading directly
to Eqs.(\ref{rel1}), (\ref{rel2}), and (\ref{rel3}).

\paragraph*{{\bf Acknowledgments.}} 
One of us (E.G.) acknowledges support from the European
Union through the Human Capital and Mobility program contract
nr. ERBCHBGCT930320.



\begin{thebibliography}{99}
\bibitem{zhuk93} M.V. Zhukov, B.V. Danilin, D.V. Fedorov, J.M. Bang,
I.J. Thompson, and J.S. Vaagen, Phys. Rep. 231 (1993) 151
\bibitem{bert93} C. Bertulani, L.F. Canto, and M.S. Hussein,
Phys. Rep. 226 (1993) 281
\bibitem{han95} P.G. Hansen, A.S. Jensen, and B. Jonson, Annu. Rev.
Nucl. Part. Sci. 45 (1995) 591
\bibitem{han87} P.G. Hansen and B. Jonson, Europhys. Lett. 4
(1987) 409
\bibitem{joh90} L. Johannsen, A.S. Jensen, and P.G. Hansen, Phys.
Lett. B244 (1990) 357
\bibitem{fed94} D.V. Fedorov, A.S. Jensen, and K. Riisager, Phys. Rev.
C49 (1994) 201
\bibitem{kob88} T. Kobayashi, O. Yamakawa, K. Omata, K. Sugimoto,
 T. Shimoda, N. Takahashi and Tanihata, Phys. Rev. Lett. 60
(1988) 2599
\bibitem{anne90} R. Anne {\it et al.}, Phys. Lett. B250 (1990) 19
\bibitem{orr92} N.A. Orr {\it et al.}, Phys. Rev. Lett.  69
(1992) 2050
\bibitem{orr95} N.A. Orr {\it et al.}, Phys. Rev. C51 (1995) 3116
\bibitem{zin95} M. Zinser {\it et al.}, Phys.~Rev. Lett. 75
(1995) 1719
\bibitem{nils95} T. Nilsson {\it et al.}, Europhys. Lett. 30 (1995) 19
\bibitem{humb95} F. Humbert {\it et al.}, Phys. Lett. B347
(1995) 198
\bibitem{zhuk94} M.V. Zhukov, L.V.Chulkov, D.V. Fedorov, B.V. Danilin,
J.M. Bang, J.S. Vaagen and I.J. Thompson,   J. Phys. G20
(1994) 201
\bibitem{kors94} A.A. Korsheninnikov and T. Kobayashi, Nucl. Phys. 
A567 (1994) 97
\bibitem{thom94} I.J. Thompson and M.V. Zhukov, Phys. Rev. C49
(1994)1904
\bibitem{zhuk95} M.V. Zhukov and B. Jonson, Nucl. Phys. A589
(1995) 1
\bibitem{bar93}, F. Barranco, E. Vigezzi, and R.A. Broglia, Phys.~Lett.
B319 (1993) 387
\bibitem{gar96} E. Garrido, D.V. Fedorov and A.S. Jensen, Phys. Rev.
 C53 (1996) 3159, and submitted for publication
\bibitem{fed94b} D.V. Fedorov, A.S. Jensen, and K. Riisager, Phys. Rev.
C50 (1994) 2372
\bibitem{gar95}  E. Garrido, D.V. Fedorov and A.S. Jensen, Phys. Rev.
C51 (1995) 3052
\bibitem{cob96}  A. Cobis, D.V. Fedorov, and A.S. Jensen, submitted 
for publication
\bibitem{newt82} R.G. Newton, Scattering Theory of Waves and Particles,
Second Edition (Springer-Verlag, New York, 1982), p.444
\bibitem{you93} B.M. Young {\it et al.}, Phys. Rev. Lett. 71 
(1993) 4124
\bibitem{tani92} I. Tanihata {\it et al.}, Phys.~Lett. B287
(1992) 307
\bibitem{you94} B.M. Young {\it et al.}, Phys. Rev. C49
(1994) 279
\bibitem{kryg93} R.A. Kryger {\it et al.}, Phys. Rev. C47
(1993) R2439
\bibitem{boh93} H.G. Bohlen {\it et al.}, Z. Phys. A344 (1993) 381

\end{thebibliography}
\end{document}